\documentclass{article}
\usepackage{frascatiphys,here,graphicx,subfigure}
\begin{document}
\title{HEAVY HADRONS IN THE RELATIVISTIC QUARK MODEL 
}
\author{D. Ebert      \\
{\em  Institut f\"ur Physik, Humboldt--Universit\"at zu Berlin, Germany} \\
R.N. Faustov and V.O. Galkin      \\
{\em Dorodnicyn Computing Centre, Russian Academy of Sciences, Russia}
}
\maketitle
\baselineskip=11.6pt
\begin{abstract}
Masses of heavy baryons and tetraquarks are calculated in the
relativistic quark model using the heavy-quark--light-diquark and
diquark-antidiquark approximations, respectively. 
\end{abstract}
\baselineskip=14pt
\section{Introduction}
Recently significant experimental progress has been achieved in heavy
hadron spectroscopy. Masses of the $\Omega_c^*$, $\Sigma_b$,
$\Sigma_b^*$ and $\Xi_b$ baryons as well as masses of several excited
charmed baryons have been measured. In the heavy meson sector several  new
states, such as $X(3872)$, $Y(4260)$, $D^*_{s0}(2317)$, $Z(4430)$ etc., were
observed which cannot be simply accommodated in the 
quark-antiquark ($q\bar q$) picture. These  states
can be considered as indications of the possible existence of
exotic multiquark states.
In this talk we briefly review our recent results for the masses of
heavy baryons and tetraquarks in  the framework of the relativistic quark
model based on the quasipotential approach in quantum chromodynamics.
We use the heavy-quark--light-diquark and diquark-antidiquark
approximations to reduce a very complicated relativistic three- and
four-body problem to the subsequent two more simple two-body
problems. The first step consists in the calculation of the masses, wave
functions and form factors of the diquarks, composed from two light
quarks or a light and heavy quark.
At the second step, a heavy baryon is treated as a
relativistic bound system of a light diquark and heavy quark. The
heavy tetraquark is considered to be a bound diquark-antidiquark system. It is
important to emphasize that we do not consider a diquark as a point
particle but explicitly take into account its structure by calculating
the form factor of the diquark-gluon interaction in terms of the diquark wave
functions.

\section{Relativistic quark model}

In the quasipotential approach the two-particle bound state with the mass
$M$ and masses of the constituents $m_{1,2}$ in momentum
representation is described by the wave
function $\Psi({\bf p})$ 
satisfying the quasipotential equation of the Schr\"odinger type
\begin{equation}
\label{quas}
{\left(\frac{b^2(M)}{2\mu_{R}}-\frac{{\bf
p}^2}{2\mu_{R}}\right)\Psi_{d,B,T}({\bf p})} =\int\frac{d^3 q}{(2\pi)^3}
 V({\bf p,q};M)\Psi_{d,B,T}({\bf q}),
\end{equation}
where 
\[
\mu_{R}=\frac{M^4-(m^2_1-m^2_2)^2}{4M^3},\quad {b^2(M) }
=\frac{[M^2-(m_1+m_2)^2][M^2-(m_1-m_2)^2]}{4M^2}.
\]
The subscript $d$ refers to the diquark, $B$ refers to the baryon
composed of a light diquark and heavy quark, and $T$ refers to the
tetraquark composed of a diquark and antidiquark.
The explicit expressions for the corresponding quasipotentials $V({\bf p,q};M)$  can be
found in Refs.\cite{hbar,tetr}.

At the first step, we calculate the masses and form factors of the
light and heavy
diquark. As it is well known, the light quarks are highly
relativistic, which makes the $v/c$ expansion inapplicable and thus,
a completely relativistic treatment of the light quark dynamics is required. To achieve this goal we closely follow our recent consideration
of the spectra of light 
mesons and adopt the same procedure to make the relativistic
potential local by replacing
$\epsilon_{1,2}(p)=\sqrt{m_{1,2}^2+{\bf p}^2}\to E_{1,2}=(M^2-m_{2,1}^2+m_{1,2}^2)/2M$. 
Solving numerically the quasipotential equation (\ref{quas}) with the
complete relativistic potential,  which depends on the
diquark mass in a complicated highly nonlinear way\cite{hbar}, we get
the diquark masses and wave functions. In order to determine the
diquark interaction with the gluon field, which 
takes into account the diquark structure, we
calculate the corresponding matrix element of the quark
current between diquark states. Such calculation leads to the
emergence of the form factor $F(r)$ entering the vertex of the
diquark-gluon interaction\cite{hbar}. This form factor is expressed
through the overlap integral of the diquark wave functions.

\section{Mass spectra of heavy baryons }

We calculated the masses of heavy baryons as the bound
states of a heavy quark and light diquark.
For  the potential of the heavy-quark--light-diquark
interaction  we used the expansion in $p/m_Q$ ($Q=c,b$). Since the
light diquark is not  heavy enough for the applicability of a $p/m_d$
expansion, it has been treated fully relativistically. The obtained
values of  masses of the ground state and excited baryons are given in 
Tables~\ref{tab:lq}-\ref{tab:xv} in comparison with available
experimental data. 

\begin{table}
\centering
\caption{\label{tab:lq}\it Masses of the $\Lambda_Q$ 
baryons (in MeV).}
\vskip 0.1 in
\begin{tabular}{ccccccc}\hline
& $Qd$ & \multicolumn{2}{c}{\hspace{-1cm}\underline{\hspace{1.cm}$Q=c$\hspace{1.cm}}}\hspace{-0.8cm}& \multicolumn{3}{c}{\hspace{-1cm}\underline{\hspace{1.8cm}$Q=b$\hspace{1.8cm}}}\hspace{-1cm}\\
$I(J^P)$& state & $M$ & $M^{\rm exp}$\cite{pdg}&  $M$ & $M^{\rm
  exp}$\cite{pdg}& $M^{\rm exp}$\cite{cdfLambda}\\
\hline
$0(\frac12^+)$& $1S$ & 2297 & 2286.46(14) & 5622 & 5624(9)&5619.7(2.4)\\
$0(\frac12^-)$& $1P$ & 2598 & 2595.4(6) & 5930 & \\
$0(\frac32^-)$& $1P$ & 2628 & 2628.1(6) & 5947 & \\
$0(\frac12^+)$& $2S$ & 2772 & 2766.6(2.4)?& 6086 & \\
$0(\frac32^+)$& $1D$ & 2874 &           & 6189 & \\
$0(\frac52^+)$& $1D$ & 2883 & 2882.5(2.2)?& 6197 & \\
$0(\frac12^-)$& $2P$ & 3017 &           & 6328 & \\
$0(\frac32^-)$& $2P$ & 3034 &           & 6337 & \\
\hline
\end{tabular}
\end{table}

\begin{table}[t,b,h]
\centering
\caption{\label{tab:sq} \it Masses of the $\Sigma_Q$ 
baryons (in MeV).}
\vskip 0.1 in
\begin{tabular}{ccccccc}\hline
&  $Qd$& \multicolumn{2}{c}{\hspace{-0.8cm}\underline{\hspace{1.2cm}$Q=c$\hspace{1.2cm}}}\hspace{-0.8cm}& \multicolumn{3}{c}{\hspace{-1cm}\underline{\hspace{2.1cm}$Q=b$\hspace{2.1cm}}}\hspace{-1cm}\\
$I(J^P)$& state & $M$ & $M^{\rm exp}$\cite{pdg,babar2940}  &  $M$ &
$M^{\rm exp}(\Sigma_b^{+})$
&
$M^{\rm exp}(\Sigma_b^{-})$\cite{cdfSigma}\\
\hline
$1(\frac12^+)$& $1S$ & 2439 & 2453.76(18)&       5805 & 5807.5(2.5)& 5815.2(2.0)\\
$1(\frac32^+)$& $1S$ & 2518 & 2518.0(5)  &       5834 & 5829.0(2.3)& 5836.7(2.5)\\
$1(\frac12^-)$& $1P$ & 2805 &            &       6122 &     \\
$1(\frac12^-)$& $1P$ & 2795 &            &       6108 &    \\
$1(\frac32^-)$& $1P$ & 2799 & 2802($^4_7$)&      6106 & \\
$1(\frac32^-)$& $1P$ & 2761 & 2766.6(2.4)?&      6076 & \\
$1(\frac52^-)$& $1P$ & 2790 &            &       6083 & \\
$1(\frac12^+)$& $2S$ & 2864 &            &       6202 & \\
$1(\frac32^+)$& $2S$ & 2912 & 2939.8(2.3)?&      6222 &\\
 $1(\frac12^+)$& $1D$ & 3014 &            &       6300 & \\
 $1(\frac32^+)$& $1D$ & 3005 &            &       6287 & \\
 $1(\frac32^+)$& $1D$ & 3010 &            &       6291 & \\
 $1(\frac52^+)$& $1D$ & 3001 &            &       6279 & \\
 $1(\frac52^+)$& $1D$ & 2960 &            &       6248 & \\
 $1(\frac72^+)$& $1D$ & 3015 &            &       6262 & \\
\hline
\end{tabular}
\end{table}

\begin{table}[t,b,h]
\centering
\caption{\label{tab:xs}\it Masses of the $\Xi_Q$ 
baryons  with the scalar diquark (in MeV).}
\vskip 0.1 in
\begin{tabular}{ccccccc}\hline
& $Qd$ & \multicolumn{3}{c}{\hspace{-1cm}\underline{\hspace{2.cm}$Q=c$\hspace{2.cm}}}\hspace{-0.8cm}& \multicolumn{2}{c}{\underline{\hspace{0.9cm}$Q=b$\hspace{0.9cm}}}\\
$I(J^P)$& state & $M$ & $M^{\rm exp}$\cite{pdg}& $M^{\rm exp}$\cite{babar} & $M$ & $M^{\rm exp}$\cite{cdfxib}\\
\hline
$\frac12(\frac12^+)$& $1S$ & 2481 & 2471.0(4)  & & 5812 &5792.9(3.0)\\
$\frac12(\frac12^-)$& $1P$ & 2801 & 2791.9(3.3)& & 6119  \\
$\frac12(\frac32^-)$& $1P$ & 2820 & 2818.2(2.1)& & 6130  \\
$\frac12(\frac12^+)$& $2S$ & 2923 &            & & 6264  \\
$\frac12(\frac32^+)$& $1D$ & 3030 &            & & 6359  \\
$\frac12(\frac52^+)$& $1D$ & 3042 &       &3054.2(1.3) & 6365  \\
$\frac12(\frac12^-)$& $2P$ & 3186 &            & & 6492  \\
$\frac12(\frac32^-)$& $2P$ & 3199 &            & & 6494  \\
\hline
\end{tabular}
\end{table}

\begin{table}[t,b,h]
\centering
\caption{\label{tab:xv}\it Masses of the $\Xi_Q$ 
baryons  with the axial vector diquark (in MeV).}
\vskip 0.1 in
\begin{tabular}{ccccccc}\hline
& $Qd$ & \multicolumn{4}{c}{\hspace{-1cm}\underline{\hspace{3.cm}$Q=c$\hspace{3.cm}}}\hspace{-0.8cm}& {\underline{\hspace{0.2cm}$Q=b$\hspace{0.2cm}}}\\
$I(J^P)$& state & $M$ & $M^{\rm exp}$\cite{pdg} & $M^{\rm exp}$\cite{belle}& $M^{\rm exp}$\cite{babar}&  $M$ \\
\hline
$\frac12(\frac12^+)$& $1S$ & 2578 & 2578.0(2.9) & & & 5937 \\
$\frac12(\frac32^+)$& $1S$ & 2654 & 2646.1(1.2) & & & 5963 \\
$\frac12(\frac12^-)$& $1P$ & 2934 &           & &   & 6249    \\
$\frac12(\frac12^-)$& $1P$ & 2928 &           & &   & 6238     \\
$\frac12(\frac32^-)$& $1P$ & 2931 &           & &   & 6237  \\
$\frac12(\frac32^-)$& $1P$ & 2900 &           & &   & 6212  \\
$\frac12(\frac52^-)$& $1P$ & 2921 &           & &   & 6218  \\
$\frac12(\frac12^+)$& $2S$ & 2984 & &2978.5(4.1)&2967.1(2.9) & 6327 \\
$\frac12(\frac32^+)$& $2S$ & 3035 &           & &   & 6341 \\
$\frac12(\frac12^+)$& $1D$ & 3132 &           & &   & 6420  \\
$\frac12(\frac32^+)$& $1D$ & 3127 &           & &   & 6410 \\
$\frac12(\frac32^+)$& $1D$ & 3131 &           & &   & 6412 \\
$\frac12(\frac52^+)$& $1D$ & 3123 &           & &3122.9(1.3) & 6403 \\
$\frac12(\frac52^+)$& $1D$ & 3087 & &3082.8(3.3)&3076.4(1.0) & 6377 \\
$\frac12(\frac72^+)$& $1D$ & 3136 &           & &   & 6390 \\
\hline
\end{tabular}
\end{table}

At present the best experimentally studied quantities are the mass
spectra of the
$\Lambda_Q$ and $\Sigma_Q$ baryons, which contain the light scalar or
axial vector diquarks, respectively. They are presented in
Tables~\ref{tab:lq}, \ref{tab:sq}. Masses of the ground
states are measured both for charmed and bottom
$\Lambda_Q$, $\Sigma_Q$ baryons. The masses of the
ground state $\Sigma_b$ and $\Sigma^*_b$ baryons were first reported
very recently by CDF\cite{cdfSigma}. 
CDF also significantly improved the accuracy of
the $\Lambda_b$ mass value\cite{cdfLambda}. For charmed baryons the masses
of several excited states are also known. It is important to emphasize that
the $J^P$ quantum numbers for most excited heavy baryons have not
been determined experimentally, but are assigned by PDG on the basis of
quark model predictions. For some excited charm
baryons such as the $\Lambda_c(2765)$, $\Lambda_c(2880)$ and
$\Lambda_c(2940)$  it is even not known if they are excitations of the
$\Lambda_c$ or $\Sigma_c$.~\footnote{In Tables~\ref{tab:lq},
  \ref{tab:sq} we mark with ? the states which interpretation is
  ambiguous.} Our calculations show that the
$\Lambda_c(2765)$ can be either the first radial (2$S$) excitation of
the $\Lambda_c$ with  $J^P=\frac12^+$ containing the light
scalar diquark or the first orbital excitation (1$P$) of the 
$\Sigma_c$  with $J^P=\frac32^-$  containing the light axial
vector diquark. The $\Lambda_c(2880)$ baryon in our model is well
described by the second orbital (1$D$) excitation of the $\Lambda_c$
with  $J^P=\frac52^+$ in 
agreement with the recent spin assignment\cite{babar2940}
based on the analysis of angular distributions in the  decays
$\Lambda_c(2880)^+\to\Sigma_c(2455)^{0,++}\pi^{+,-}$. Our model
suggests that the   charmed baryon $\Lambda_c(2940)$, recently
discovered  by BaBar and 
confirmed by Belle\cite{babar2940}, could be the first radial (2$S$) excitation of
the $\Sigma_c$ with $J^P=\frac32^+$ which mass is predicted slightly
below the experimental value. If this state  proves to be an excited
$\Lambda_c$, for which we have no candidates around 2940 MeV, then it
will indicate that excitations inside the diquark should be also
considered.~\footnote{The $\Lambda_c$ baryon with the first
orbital excitation of the diquark is expected to have a mass in this
region.}   The $\Sigma_c(2800)$ baryon can be identified in our model
with one of the orbital (1$P$) excitations of the $\Sigma_c$
with $J^P=\frac12^-, \frac32^-$ or $\frac52^-$ which predicted mass
differences are  less than 15 MeV. Thus masses of all these states
are  compatible with the experimental value within errors.       
  
Mass spectra of the $\Xi_Q$ baryons with the scalar and axial vector light
($qs$) diquarks are given in Tables~\ref{tab:xs}, \ref{tab:xv}. Experimental
data here are available mostly for charm-strange baryons. We can identify the
$\Xi_c(2790)$ and $\Xi_c(2815)$ with the first orbital (1$P$)
excitations of the $\Xi_c$  with $J^P=\frac12^-$ and $J^P=\frac32^-$,
respectively, containing the light scalar diquark, which is
in agreement with the PDG\cite{pdg} assignment. Recently
Belle\cite{belle} reported the first observation of two baryons 
$\Xi_{cx}(2980)$ and $\Xi_{cx}(3077)$, which existence was also 
confirmed by BaBar\cite{babar}. The $\Xi_{cx}(2980)$ can be interpreted
in our model as the first radial (2$S$) excitation of the $\Xi_c$  with
$J^P=\frac12^+$ containing the light axial vector diquark. On the other hand the
$\Xi_{cx}(3077)$  corresponds to the second orbital (1$D$) excitation
in this system with $J^P=\frac52^+$.  The new charmed baryons
$\Xi_c(3055)$ and $\Xi_c(3123)$, very recently announced by
BaBar\cite{babarxic}  can be interpreted in our
model as the second orbital ($1D$) excitations of the $\Xi_c$  with
$J^P=\frac52^+$ containing scalar and axial vector diquarks,
respectively. Few months ago the D0 Collaboration reported
the discovery of the $\Xi_b^-$ baryon. The CDF
Collaboration\cite{cdfxib} confirmed this observation and gave the
more 
precise value of its mass. Our model prediction is in a reasonable
agreement with these new data.   

\section{Masses of  heavy tetraquarks}

\begin{table}
  \caption{\it Comparison of theoretical predictions for  the masses of
   charm diquark-antidiquark states $cq\bar c\bar q$ (in MeV) and
   possible experimental candidates.}
  \label{tab:cemass}
\vskip 0.1 in
\centerline{
\begin{tabular}{cccccc}
\hline
State&
\multicolumn{3}{l}{\underline{\hspace{2.6cm}Theory\hspace{2.6cm}}}
\hspace{-1.5cm}& 
\multicolumn{2}{l}{\underline{\hspace{0.9cm}Experiment
    \hspace{0.9cm}}} 
\hspace{-1.5cm}  \\
$J^{PC}$ &EFG & Maiani et al. &Maiani et al.($cs\bar c\bar s$) 
&state& mass\\
\hline
$1S$\\
$0^{++}$ & 3812 & 3723& & &\\
$1^{++}$ & 3871& 3872$^\dag$& &$X(3872)$ &$3871.9(0.5)$ \\
$1^{+-}$ & 3871& 3754&& &\\
$0^{++}$& 3852 & 3832&& &\\
$1^{+-}$& 3890 & 3882&& &\\
$2^{++}$& 3968 & 3952&&$Y(3943)$&$3943(11)(13)$ \\
$1P$\\
$1^{--}$&4244 & &$4330(70)$&$Y(4260)$ &$4259(8)(^{2}_{6})$\\
\hline
 \end{tabular}}
\vspace*{-0.4cm}
\flushleft{\hspace*{0.5cm}${}^\dag$ input}
\end{table}
To calculate the masses of heavy tetraquarks 
we considered them as the
bound states of a heavy diquark and antidiquark. 
In Table~\ref{tab:cemass} we compare our results (EFG\cite{tetr}) for
the charm 
diquark-antidiquark bound states with the predictions of Ref.\cite{mppr}. 
The differences in some of the
mass values can be attributed to the substantial distinctions in
the used approaches. We describe the diquarks dynamically as
quark-quark bound 
systems and  calculate their masses and form factors, while in
Ref.\cite{mppr}  they
are treated only phenomenologically. Then we consider the tetraquark
as purely the 
diquark-antidiquark bound system.  In distinction Maini et al. 
consider a
hyperfine interaction between all quarks  which, e.g., causes the
splitting of $1^{++}$ and $1^{+-}$ states arising from the $SA$
diquark-antidiquark compositions.  
From Table~\ref{tab:cemass}, where we also give possible experimental
candidates for the neutral tetraquarks with hidden charm,  we see that our
calculation supports  the assumption\cite{mppr} that $X(3872)$ can be
the axial vector 
$1^{++}$ tetraquark state composed from the scalar and axial vector
diquark and antidiquark in the relative $S$ state.  
On the other hand, in our model 
the lightest scalar $0^{++}$ tetraquark is
predicted to be above the open charm threshold $D\bar D$
and thus to be broad, 
while in the model\cite{mppr} it lies 
few MeV below this threshold, and thus is predicted to be narrow. Our
$2^{++}$ tetraquark also lies higher than the one in Ref.\cite{mppr}. We
find that $Y(4260)$ cannot be interpreted as $P$ state $1^{--}$
of charm-strange diquark-antidiquark, since its mass is found to
be $\sim 200$ MeV higher. A more natural
tetraquark interpretation could be the $P$ state $([cq]_{S=0}[\bar
c\bar q]_{S=0})$  which mass is predicted in our
model to be 
close to the mass of  $Y(4260)$ (see Table~\ref{tab:cemass}). 
Then the $Y(4260)$ would decay dominantly into $D\bar D$ pairs.

\section{Conclusions}

We found that presently available experimental data for the masses of
the ground and excited 
states of heavy baryons can be accommodated in the picture
treating a heavy baryon as the bound system of the light diquark
and heavy quark, experiencing orbital and radial excitations 
between these constituents.  
 It was argued that the $X(3872)$ and $Y(4260)$
can be the neutral charm tetraquark states. If they are really  tetraquarks,
one more neutral and two charged tetraquark states should exist with close
masses. 

This work was supported in part by the {\it Deutsche
Forschungsgemeinschaft} under contract Eb 139/2-4 and by the {\it Russian
Foundation for Basic Research} under Grant No.05-02-16243. 

\end{document}